\documentclass[journal=jacsat,manuscript=article]{achemso}

\usepackage[version=3]{mhchem} 
\usepackage{graphicx}
\usepackage{xcolor}
\usepackage[numbers]{natbib}

\definecolor{darkgreen}{rgb}{0,0.4,0}

\usepackage{color}
\definecolor {myc} {rgb} {0,0,0}  

\author{M. Milani$^{1}$}
\author{K. Ahmad$^1$}
\author{E. Cavalletti$^1$}
\author{C. Ligoure$^1$}
\author{L. Cipelletti$^{1,2}$}

\author{M. Kongkaew$^3$}
\author{P. Trens$^3$}

\author{L. Ramos$^1$}
\email{laurence.ramos@umontpellier.fr}

\affiliation{%
$^1$Laboratoire Charles Coulomb (L2C), Universit\'e Montpellier, CNRS, Montpellier, France\\
$^2$Institut Universitaire de France, Paris, France\\
$^3$ Laboratoire ICGM, Universit\'e Montpellier, CNRS, Montpellier, France
}%

\title[An \textsf{achemso} demo]
  {Synthesis of spherical mesoporous silica beads with tunable size, stiffness and porosity}


\begin{document}



\begin{abstract}
We present an innovative template-free
water-based sol-gel method to produce uniform mesoporous silica beads of millimeter size, which have tunable size, stiffness and porosity, and  could be used for adsorption applications. Our protocol exploits an in-situ enzymatic reaction to produce spherical beads of hydrogel from a charge-stabilized suspension of silica nanoparticles confined in a millimetric drop suspended in a non-miscible oil. Once the gelation step is complete, the spherical bead of gel is cleaned from oil and deposited onto a hydrophobic surface and let dry. Separating the gelation to the drying steps ensures a spatially uniform gel and allows us to perform a solvent exchange before drying. 
For all beads, we observe a crack-free drying process leading to the formation of stiff quasi-spherical beads with diameter in the range 1 to 5 mm and Young modulus in the range $(0.1-2)$ GPa and narrow pore size distribution, centered around $10$ to $25$ nm depending on the experimental conditions. Finally, to demonstrate the potentiality of these materials, we graft on the bead surface aminosilane molecules, and quantify their CO$_2$ adsorption efficiency. Overall, the production method we have developed  is simple, readily adaptable, and offers promising materials for adsorption, storage, catalysis and chromatography.
\end{abstract}


\section{Introduction}

Silica monoliths are key for several different applications such as adsorption,\cite{tao2011superwetting,sun2011hierarchically,rodrigues2013strategies,galarneau2016hierarchical} catalysis,\cite{galarneau2016hierarchical,ren2012porous} and chromatography \cite{galarneau2006spherical,zhong2008direct,wang2012facile,nuzhdin2016hkust,meinusch2015synthesis}, due to their large specific surface area and porosity, small pore size ($2-50$ nm), low effective thermal conductivity and low dielectric constant\cite{dorcheh2008silica}. In contrast to silica powders or films, producing silica monoliths by drying a hydrogel formed through a sol-gel method is a complicated task. In fact, while drying, a hydrogel is exposed to capillary and adhesion forces\cite{brinker2013sol,thiery2015water,thiery2016drying}. Both forces impose enormous stresses on the silica network, eventually leading to cracking, delamination, and crumbling. 
An approach to overcome this problem is supercritical drying\cite{brinker2013sol,estella2008effect}. However, supercritical drying involves expensive high-pressure equipment, making this approach not easily accessible. For this reason, other methods exploiting ambient drying have been adopted.  

Mesoporous silica monoliths obtained under ambient drying conditions can be synthesized by using a secondary material to scaffold a silica network~\cite{chen2012one,grund2006monolithic,guillemot2013percolation,zhang2023fumed}. Scaffolding exploits secondary materials such as carbon, polymers, and cellulose that stiffen the network and prevent cracking during drying. An alternative strategy for the production of mesoporous silica monoliths consists in introducing flexibility in the silica network~\cite{husing1998aerogels,hwang2008optimization,zu2018transparent}.  Modifying silica gels with a silylating agent allows in particular the solid matrix to swell back once the gel is completely dried, resulting in a hydrophobic aerogel. However, despite being successful, producing silica monoliths by scaffolding or by introducing flexibility is complicated and costly.

Recently, Marszewski et al. produced mesoporous silica slabs by drying a colloidal suspension poured onto a non-adhesive Teflon mold covered with a perfluorochemical  liquid~\cite{marszewski_thick_2019, marszewski_transparent_2022}. Using these hydrophobic surfaces, and ambient drying conditions, the authors succeeded in obtaining, crack-free mesoporous cylindrical monoliths of diameter several centimeters and of thickness a few millimeters. The mesoporous slabs were produced without any chemical modification of the silica nanoparticles, making their method easily replicable. However, in this work, the gelation and drying steps occur concomitantly, impeding a full control of the resulting mesostructures. Indeed, as demonstrated in a recent study~\cite{nmar_structuration_2023}, the coupling between gelation and drying may lead to heterogeneous gels. Hence, mastering independently and sequentially the gelation and the drying processes would allow one to better control the mechanical and textural properties of the fully dried materials. Moreover, when gelation and drying occur concomitantly, a solvent exchange process, is not feasible, whereas several studies acknowledge that solvent exchange may improve the textural properties of the materials~\cite{smith1992preparation,land2001processing,kim2003synthesis}. Indeed, by exchanging the water with a solvent with lower surface tension, it is possible to minimize the capillary forces and consequently the failure events during drying~\cite{smith1992preparation,land2001processing,kim2003synthesis}.

\begin{figure}
\includegraphics[width=\columnwidth]{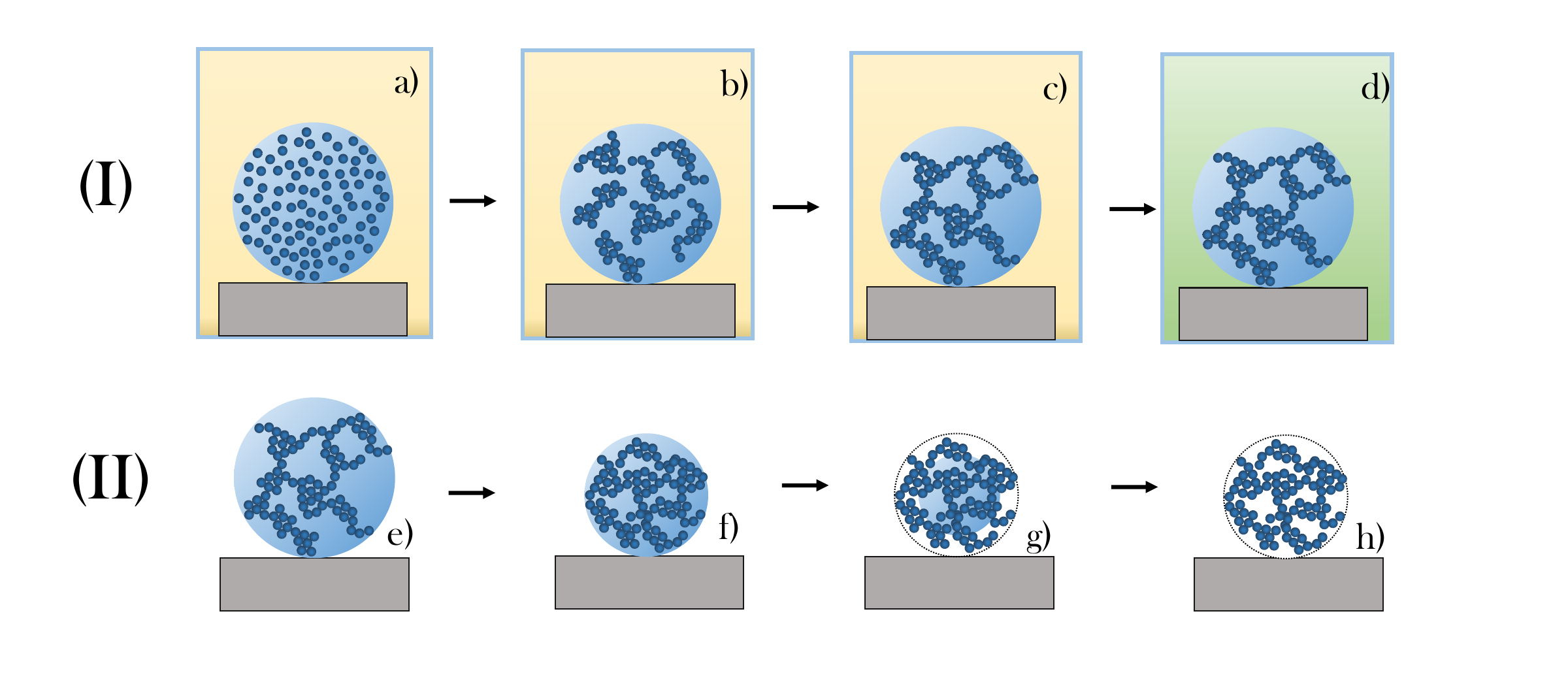}
\caption{Sketch of the two-step process to produce mesoporous silica beads with controlled characteristic sizes: the first step (I) consists in the preparation of homogeneous spherical beads of colloidal gel, and the second step (II) is the drying of the bead of gel. a) Immediately after mixing, a drop of silica nanoparticle (NP) aqueous suspension is transferred on a hydrophobic surface immersed in a bath of immiscible oil. b) A slow in-situ enzymatic reaction releases ions into the suspension, which leads to the formation of fractal clusters due to NPs aggregation. c) A colloidal gel is obtained when the clusters interconnect forming a volume spanning network. d) Once the gelation is complete, it is possible to immerse the bead in an ethanol bath to exchange water with ethanol (This step is optional for the preparation). e) \textcolor{myc}{Once the bead is removed from the oil (or ethanol) and exposed to air, the the drying process starts. f) The bead of gel shrinks homogeneously due to solvent evaporation. During this step the NPs density increases with time. g) At a certain time the bead stops shrinking and air invades the sample. This occurs before the NPs are highly compacted leading to the formation of mesopores characterizing the dried beads of panel h).}}
\label{Fig_synthesis}
\end{figure}

In this work we produce mesoporous silica beads with controlled characteristic sizes and stiffness, by using the two-step process sketched in Fig.~\ref{Fig_synthesis}. We first produce homogeneous beads of colloidal gel and we then let them dry in ambient conditions. For the gelation step, we first immerse in an immiscible oil a drop of silica nanoparticles (NPs) dispersed in an aqueous solvent. Then, thanks to an enzymatic reaction taking place in the solvent of the NPs, we induce in-situ the destabilization of the NPs to yield a fractal colloidal gel. As we will show, the in-situ destabilization of the NPs to form a gel and the clear separation between the gelation step and the drying step are key for the production of fully homogeneous materials with controlled textural properties. We provide simple means to change the textural and mechanical properties of the beads by playing with the initial volume fraction of the NPs and their size. Moreover, we also demonstrate that the textural properties of the mesoporous beads can be improved by exchanging water with ethanol. Finally, to illustrate the potentiality of our mesoporous beads for applications, we graft the beads with aminosilane molecules and probe the capability of the chemically modified beads to capture carbon dioxide.

\section{Experimental section}

\subsection{Materials}

We use commercial suspensions of charge-stabilized silica nanoparticles (NPs),  Ludox SM-30 and Ludox TM-50, from Sigma-Aldrich. The NPs are initially suspended in water and have a nominal diameter $a=6$ nm, respectively $a=25$ nm. We prepare NP suspensions at different initial volume fraction $\varphi_0$ in the range $(1.5-15)$ \%, in an aqueous solvent comprising urea (at concentration $1$ M) and the enzyme urease (U 5378) from Canavalia ensiformis, Sigma-Aldrich, at a fixed concentration ($m_e\approx 30$ U/mL). We exploit the urea-urease enzymatic reaction, CO(NH$_2$)$_2$ + 2H$_2$O $\rightarrow$ NH$_4^+$ + NH$_3$ + HCO$_3^-$, to release in-situ ions, thus screening the repulsive electrostatic potential between NPs and triggering gel formation through NP aggregation~\cite{wyss2004small}.
For the oil bath we use a silicon oil (47 V 100 from VWR chemicals). We saturate the oil with water such that no water diffusion occurs during the gelation process and as long as the bead remains immersed in oil (Fig.~\ref{Fig_synthesis}a-c), thus fully preventing water from evaporating from the bead. The oil has a density $\rho=0.97$g/mL at $25^{\circ}$C and the water/oil  interfacial tension is $\gamma=40$ mN/m. For the grafting of the dried beads, we use toluene (99.99$\%$, Sigma Aldrich), and 3-aminopropyl(triethoxysilane) (APTES) (99.9$\%$, Sigma Aldrich).

To prepare the hydrophobic surface onto which the drop is deposited, we use trimethoxy-(octadecyl)-silane (Sigma Aldrich).  Typically, a water/ethanol solution ($50\%/50\%$) of trimethoxy-(octadecyl)-silane ($20$ mg/mL) is deposited on a glass slide and heated in an oven ($1$ h ramp from $30^{\circ}$C to $70^{\circ}$C, and then $12$ h at $70^{\circ}$C). The contact angle formed by an aqueous drop ($30-40$ $\mu$L) deposited on such surface is approximately $120^{\circ}$.

\subsection{Beads production}

The beads production is based on the protocol sketched in Fig.~\ref{Fig_synthesis}. 
A drop of silica nanoparticle (NP) suspension comprising urea and urease is deposited on an hydrophobic surface in a silicon oil bath. The oil density is smaller than that of the suspension so that the drop remains still on the surface. The suspensions are prepared at different volume fractions $\varphi_0$, by diluting the initial Ludox suspension with a mixture of milli-Q water, urea, and the appropriate amount of urease. 
With time, the enzymatic reaction triggers the NPs aggregation leading to gelation. The gelation time slightly varies with $\varphi_0$ and is typically of the order of $3000$ s, as probed in situ using the dynamic light scattering apparatus described in Ref.~\citet{milani2024space}. The mechanical and structural properties of colloidal gel evolve in time, mainly during to the first hours after gelation~\cite{cipelletti2000universal}. Thus, we let the bead in oil for at least $24$ h to ensure that the gelation and aging processes are complete, but the duration of the immersion in oil is not a crucial parameter and the bead can stay much longer in oil, without any impact on the sample properties.
The bead is then removed from the oil bath and rinsed thoroughly by plunging it three times in a cyclohexane bath. We also produce beads with ethanol instead of water, using a solvent exchange step: to do so, we immerse the beads of gel prepared using the standard protocol with water as solvent into an ethanol bath for at least four days, thus allowing ethanol to completely replace water within the gel. For this step the amount of ethanol in the bath is much larger than the solvent within the bead. The bead of colloidal gel, with either water or ethanol as the solvent, is then let to fully dry in open air, typically at $35^{\circ}$C for $12$ h. From our experiments, we anticipate that the drying temperature is not a crucial parameter: we have indeed checked that by varying the drying rate we do not alter the mechanical or textural properties of the silica beads.
Typically, we use drops of volume $30$ $\mu$L yielding beads of gel of diameter $d_0=3.8$ mm, and dried beads of diameter $d$ in the range $(1.4-2.6)$ mm. In addition, we prepare gel beads with different sizes, by varying the initial volume of the drop suspensions, from $4$ $\mu$L to $1.4$ mL.

\subsection{Beads grafting}

The grafting procedure is based on the protocol described in Ref.~\citet{cueto2021aptes}. In brief, a total of $60$ mg of dried silica beads are first suspended in $2.5$ mL of toluene. After that, 100 $\mu$L (0.43 mmol) of 3-aminopropyl(triethoxysilane) (APTES, 99.9$\%$ from Sigma-Aldrich) are added to the silica bead suspension. The reaction mixture is then heated up to $50^{\circ}$C and kept under stirring for $24$ h. The obtained product is finally washed with ethanol and dried at $50^{\circ}$C in air.

\subsection{Characterization Methods}


\subsubsection{Dynamic light scattering}

We use dynamic light scattering (DLS) to follow the gelation process and check the uniformity of the gelation process within the drop, in order to ensure that the resulting gel bead is spatially uniform. We use an ad-hoc apparatus that allows us to measure the microscopic dynamics with space- and time-resolution within our millimetric spherical samples~\cite{milani2024space}. Thanks to this set-up, we follow in-situ the sample gelation at different radial positions along a drop diameter, from the edge of the drop to the center. To do so, we probe the quiescent dynamics of the sample, by measuring a correlation index $c_I$, which quantifies the microscopic dynamics as inferred by the temporal fluctuations of the speckle pattern formed by light scattered by the sample~\cite{duri2005time}:
\begin{equation}
    c_{I}(t,\tau,x) = A \bigg( \frac{\langle I_p(t)I_p(t+\tau) \rangle_{x} }{\langle I_p(t)\rangle_{x}\langle I_p(t+\tau_0) \rangle_{x}}-1\bigg)
\label{eq:ci_correlation_function_PCI}
\end{equation}
with $\tau$ the time delay between images of the speckle pattern, $I_p(t)$ the scattered intensity of the $p$-th pixel of a region of interest (ROI) of the speckle image centered around position $x$ along the bead diameter at time $t$, $\langle ...\rangle_x$ a spacial average over all pixels in a ROI centered in $x$ and $A$ a normalization constant. Typically the size of the ROI along the drop diameter is $200$ $\mu$m, setting the coarse graining spatial resolution. The typical length scale over which the microscopic dynamics are measured is set by the scattering angle, which is here $\theta\approx 90^{\circ}$, corresponding to a length scale of $45$ nm. 

The correlation index is a robust indicator to characterize time-dependent phenomena such as gelation and aging of gels~\cite{cipelletti2000universal}, spatial heterogeneities in rearranging foams~\cite{sessoms2010unexpected}, and many others processes. In this work, we probe the gelation process by following the time evolution of the correlation index for the different ROIs for a fixed delay $\tau=0.5$ ms.

\subsubsection{Small-angle X-ray scattering}

To probe the microscopic structure of beads of colloidal gels we use small-angle X-ray scattering (SAXS). Experiments have been performed at the beam line Swing of the Synchrotron Soleil (France). To run the measurements, we remove the bead of gel from the oil, rinse it  and place it immediately onto a hydrophobic surface enclosed in a Teflon box with two kapton windows in order to minimize sample evaporation. Measurements are taken roughly $2$ min after the removal of the bead from the oil bath. For comparison, SAXS measurements are also performed on colloidal gels prepared directly in sealed capillaries (quartz capillaries of diameter $1.5$ mm).  To prepare the gels in the capillaries we use the same preparation as the one used for the beads, but instead of forming a drop in an oil bath, we inject a few microliters of the suspension in a capillary, which we immediately seal, and let the sample gel in-situ. SAXS data are analyzed using standard procedures for background subtraction and azimuthal averaging of the 2D-scattering pattern.

\subsubsection{Rheology}

To measure the mechanical properties of a porous silica bead, we place the bead in between the two flat glass plates of a rheometer (MRC502 by Anton Paar), and we impose a uniaxial compression, at a fixed compression rate $\dot{\varepsilon}$, defined as $\dot{\varepsilon}=\dot{h}/h$,  where $h$ is the plate-to-plate distance. We use a compression rate $\dot{\varepsilon}=0.1$ s$^{-1}$, but we have checked that, for $\dot{\varepsilon}$ in the range $(10^{-3}-1)$ s$^{-1}$, the Young modulus is independent of the applied strain rate, as expected from a linear elastic measurement.

The rheometer records the normal force $F_n$ as a function of the distance $h$. We obtain the Young modulus of the bead $E$, by using the Hertz contact model~\cite{johnson1982one}: $F_n = \frac{4}{3} \frac{E}{1-\nu^{2}} \bigg(\frac{d_0}{2}\bigg)^{1/2} d_c^{3/2}$. Here $d_c=d_0-h$, with $d_0$ the diameter of the bead, is the indentation depth. The Poisson ratio $\nu$ is assumed to be $\nu$=0.5, as expected for an uncompressible material. 

\subsubsection{Adsorption techniques}

The textural properties of the different fully dried beads are determined by nitrogen sorption at $77$ K using a 3Flex device (Micromeritics, USA). The beads are first degassed at $473$ K under secondary vacuum for $12$ h before analysis. The specific surface area is determined using the Brunauer–Emmett–Teller (BET) transform of the sorption isotherms in the range $0.05<p/p^{\circ}<0.25$, where $p$ and $p^{\circ}$ are the equilibrium pressure and the saturation vapour pressure of nitrogen at $77$ K respectively, taking $0.162$ nm$^{2}$ as cross-sectional area for nitrogen. The absence of microporosity is determined by a t-plot, using the Harkins and Jura equation as a reference sorption isotherm~\cite{harkins1944surfaces}. The pore size distributions are obtained by using the Barrett–Joyner–Halenda (BJH) model on the adsorption branches of the sorption isotherms\cite{rouquerol2013adsorption,bardestani2019experimental}.
The carbon dioxide sorption isotherm is obtained using the same apparatus, following the same degassing procedure($12$ h under secondary vacuum at $473$ K). We use ultra-pure CO$_2$ gas as provided by Linde. The sorption experiments are performed at $25^{\circ}$C, using a ISO sub-ambient temperature controller. When heating the beads at $473$ K during the degassing step we observe that they turn into a brown color, indicating the presence of some residual oil. Analysis of thermogravimetric measurements show however that the amounts of oil is less than 5$\%$ of the dried mass (data not shown). Moreover, a stronger confirmation of the fact that the remaining oil does not prevent the functionality of the beads lies in the  successful grafting of APTES onto the beads.

\subsubsection{IR Spectroscopy and NMR}
The APTES grafting efficiency is checked by infrared spectroscopy and solid state NMR.
The infrared spectra are obtained at ambient temperature using a Bruker Tensor 27 spectrometer equipped with a KBr beam splitter, a Black body source, a DTGS detector and an attenuated total reflectance (Znse ATR) device. $^{29}$Si and $^{13}$C solid-state NMR spectra are obtained using cross-polarization and magic-angle spinning techniques (CP-MAS) on a Bruker DSX 300 MHz spectrometer.

\section{Results}

\subsection{From a liquid drop to an homogeneous hydrogel bead}

\begin{figure}
\includegraphics[width=\columnwidth]{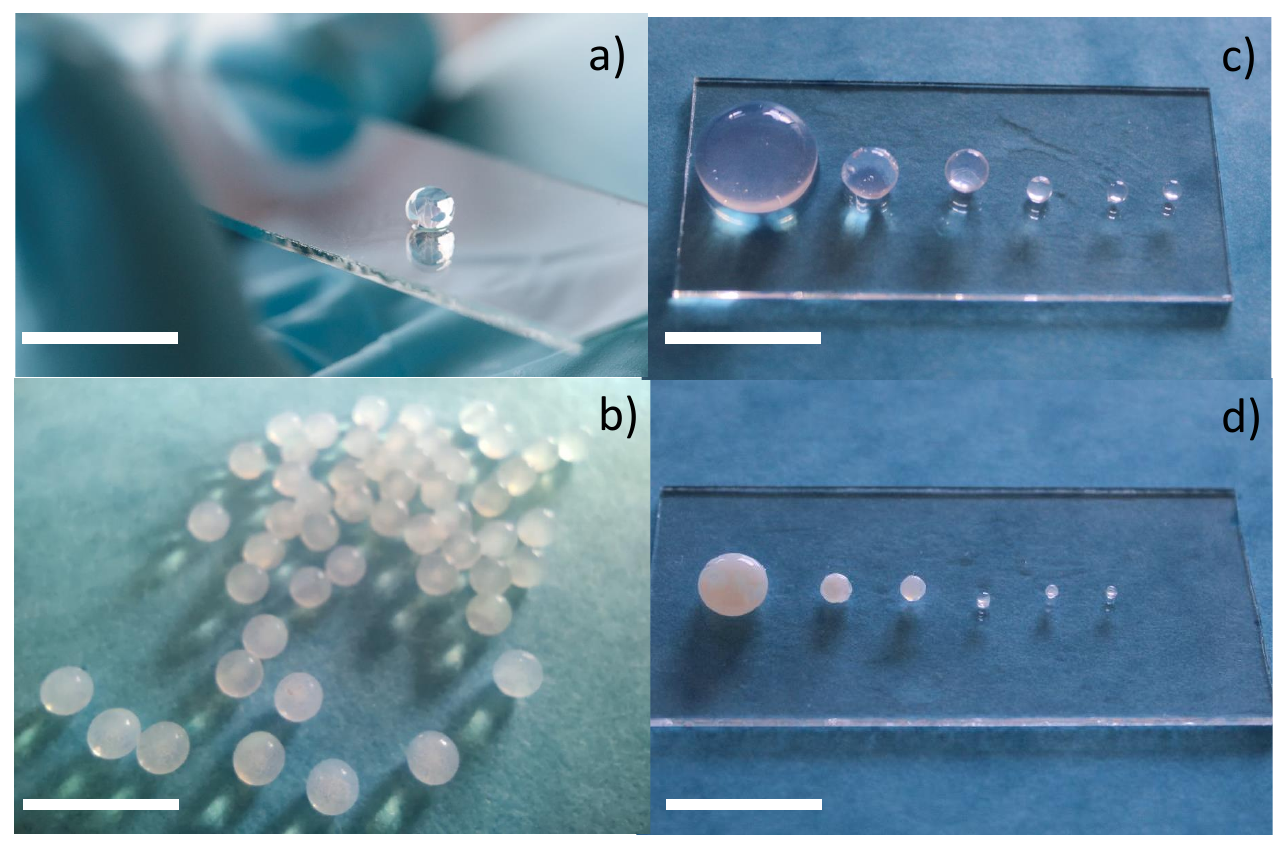}
\caption{a) Image of a bead of colloidal gel prepared with an initial volume fraction $\varphi_0=5\%$ (with NPs of diameter $6$ nm). The picture is taken right after extracting the bead from the oil bath. The scale bar represents $1$ cm. Credits: Mélanie Challe / CNRS Images. b) Mesoporous silica beads obtained from the drying (at $35^{\circ}$C for $24$ h) of beads similar to the one shown in a). The scale bar represents $0.7$ cm. c) Image of beads of colloidal gel prepared with an initial volume fraction $\varphi_0=5\%$ (with NPs of diameter $6$ nm) and different volumes: from $4$ $\mu$L up to $1.4$ mL, from right to left. The beads are nearly spherical, except the  largest one. The scale bar represents $1.3$ cm. d) Mesoporous silica beads obtained from the drying at $35^{\circ}$C for $24$ h of the beads shown in c). The scale bar represents $1.3$ cm.}
\label{Fig_imaging_results}
\end{figure}

We show in Fig.~\ref{Fig_imaging_results} a bead of colloidal gel prepared from a drop (volume $30$ $\mu$L) of suspension of NPs of radius $a=6$ nm with an initial volume fraction $\varphi_0=5\%$. The bead is almost a perfect sphere. Very generally, the deposition of a small liquid drop in an oil bath on the top of a hydrophobic surface allows one to obtain a very regular spherical shape. Indeed, the low density mismatch between the oil and the drop ($\Delta\rho = 0.02$ g/mL) minimizes the flattening of the drop due to gravity, while the use of the hydrophobic surface minimizes the contact area between the drop and the flat surface. In these conditions, the shape of a liquid drop is uniquely controlled by the capillary length, $l_c$, and drops of size smaller than $l_c$ are expected to be spherical. The capillary length reads $l_c=\sqrt{\frac{\gamma}{\Delta\rho g}}\approx 13$ mm, with $\gamma=40$ mN/m the interfacial tension between water and silicon oil. Consistently, we find that gel beads smaller than $\sim 6.5$ mm are nearly spherical while a bead of volume $\frac{\pi l_c^3}{8}=1.4$ mL is flattened by gravity (left bead in Fig.~\ref{Fig_imaging_results}c). Once the gelation processed is completed, the elastic modulus of the gel is high enough to avoid the collapse of the bead under its own weight.

\begin{figure}
\includegraphics[width=0.5\columnwidth]{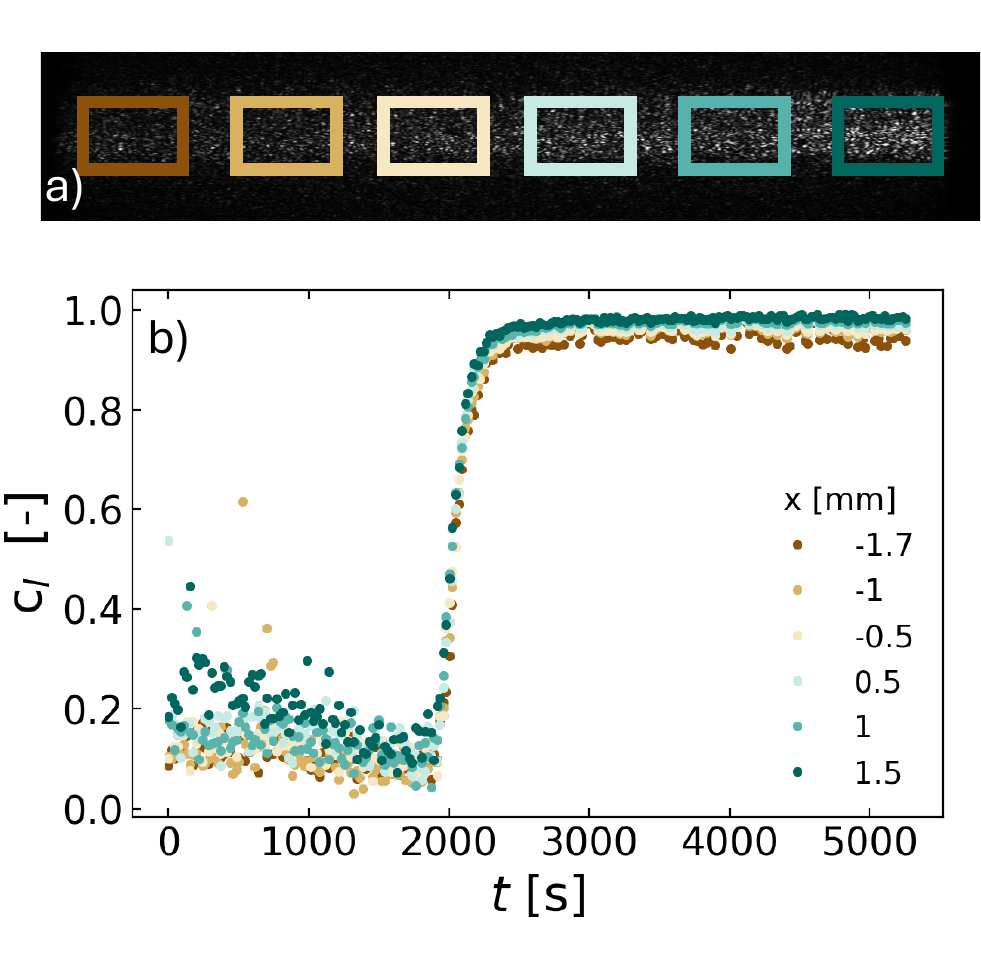}
\caption{a) Image of a speckle pattern formed by a laser beam that illuminates from the right a bead of gel along its diameter. The gel is prepared at an initial volume fraction $\varphi_0=5\%$ (with NPs of diameter $6$ nm). Data points are color-coded according to the ROIs shown in a). The bead is surrounded by oil, the colored rectangles represent the ROIs of length $200$ $\mu$m. b) Correlation index $c_I$ as a function of time $t$ for a colloidal sample prepared at $\varphi_0=10\%$, with NPs of diameter $6$ nm.}
\label{Fig_charaterization}
\end{figure}

\subsubsection{Microscopic dynamics}

We use DLS to follow the gelation process inside the drop and check the homogeneity of the beads thus formed. Figure~\ref{Fig_charaterization}a shows the speckle pattern produced by the laser when propagating along the diameter of a gel bead. Using Eq.\ref{eq:ci_correlation_function_PCI}, we measure the microscopic dynamics of the nanoparticles confined in the drop for different regions of interest, ROIs (highlighted by the colored rectangles) along a drop diameter.  We plot in Fig.~$\ref{Fig_charaterization}$b the time evolution of the correlation index $c_I( t,\tau,x)$ computed for different positions along the drop diameter, with $\tau=5$ ms. Time $0$ corresponds to the beginning of the measurement, which typically starts $2$ min after adding urease to a suspension of NPs in a mixture of urea and water. At short times, $c_I( t)\approx0$, reflecting the fact that the sample dynamics are very fast and correspond to a fluid-like sample, constituted by NPs and/or NPs aggregates which undergo Brownian motion. At time $t=t_g\sim2000$ s, $c_I$ increases sharply, reflecting the abrupt slowing down of the dynamics. This slowing down is due to the aggregation of the NPs clusters into a system-spanning network~\cite{carpineti1993transition}. Thus, we identify $t_g$ as the gelation time. 
Interestingly, we find that data acquired at different positions (from the drop center, $x=0$ to the drop edges, $x \sim \pm 1.8$ mm) almost perfectly overlap, indicating that the gelation occurs essentially at the same time throughout the drop.  This finding is a strong indication that the gels are spatially homogeneous and should be contrasted to the case of drops not immersed in oil, for which gelation competes with drying. For the latter, using the same DLS set-up, we have measured that the concomitant occurrence of gelation and drying processes leads to a significantly shorter gelation time in the core than at the edge of the drop. Hence, our experiments demonstrate that a clear decoupling between the gelation step and the drying steps, as ensured by the immersion of the drop in an oil bath that prevents solvent evaporation, is crucial to produce spatially homogeneous beads of colloidal gel.

\subsubsection{Gel microstructure}
\begin{figure}
\includegraphics[width=0.6\columnwidth]{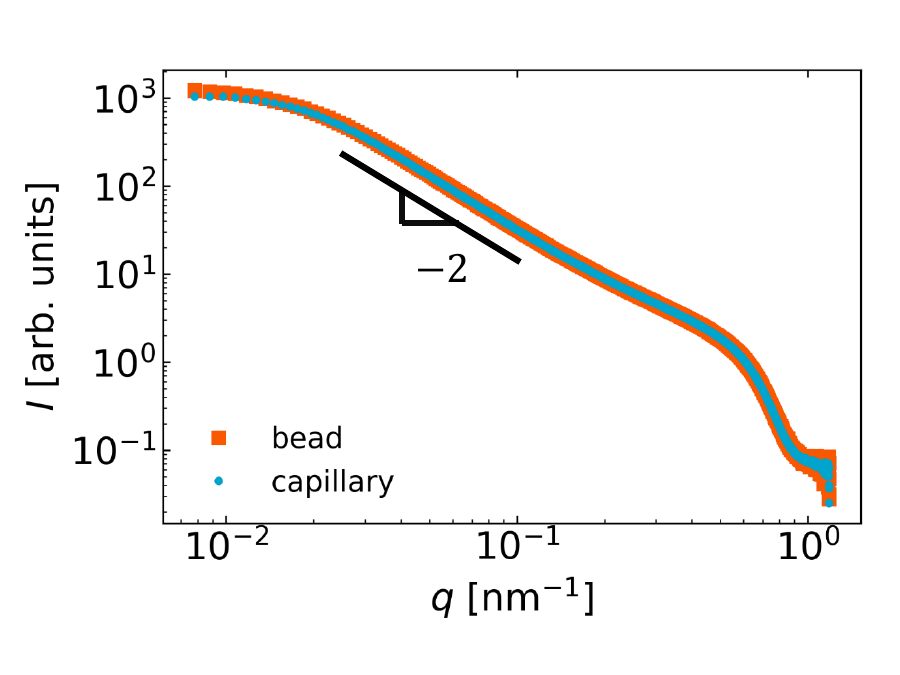}
\caption{Comparison of the scattered intensity of a bead of gel (blue dots) and a bulk gel prepared in a capillary (orange squares). Colloidal gels are prepared with an initial volume fraction $\varphi_0=2.3\%$ (with NPs of diameter $6$ nm). Scattering data have been normalized to overlap at large $q$.}
\label{Fig_dynamics_and_xray}
\end{figure}

The gel microstructure is characterized by small-angle X-ray scattering. Very generally, we find that the scattered intensity of two gels prepared in a drop immersed in oil or in a glass capillary perfectly overlap over the whole accessible range of scattering vector $q$. As an illustration, we report in Fig.\ref{Fig_dynamics_and_xray} the data for gels prepared with NPs of diameter $6$ nm at a volume fraction $\varphi_0=2.3\%$. The scattering curves are typical of fractal colloidal gels~\cite{carpineti1993transition,carpineti1995mass}: at high $q$, $q>0.5$ nm$^{-1}$, which corresponds to the smallest length scales, the scattered intensity correspond to the signal of individual NPs. At intermediate $q$, the scattered intensity decays as a power law with the  scattering vector, $I \sim q^{-d_f}$, where $d_f$ is the gel fractal dimension. A fit to the data of a power law yields $d_f=2.0\pm0.1$. This value is in good agreement with the findings of previous works for similar colloidal gels~\cite{manley_time-dependent_2005, aime2018power}. 
The scattering intensity eventually plateaus at small $q$, indicating that the gel structure is homogeneous at sufficiently large length scale, signing the end of the fractal regime. The cross-over occurs at a characteristic scattering vector $q^{*}\approx0.02$ nm$^{-1}$, from which we extract the characteristic cluster size of the gel as $\zeta=2\pi/q^{*}$. We find $\zeta\approx300$ nm for the gel shown in Fig.~\ref{Fig_dynamics_and_xray}.

\subsection{From an hydrogel bead to a mesoporous silica bead}

\subsubsection{Imaging}

The second step to prepare the mesoporous silica beads is the drying stage. Figure~\ref{Fig_imaging_results}b displays an assembly of
silica beads obtained by drying beads of gel following the protocol described above. The beads exhibit a very narrow size dispersity, demonstrating the good control of the bead size reached with our protocol, while the size of the beads can be easily tuned by varying the size of the beads of gel (Fig.~\ref{Fig_imaging_results}d).

\begin{figure}
\includegraphics[width=\columnwidth]{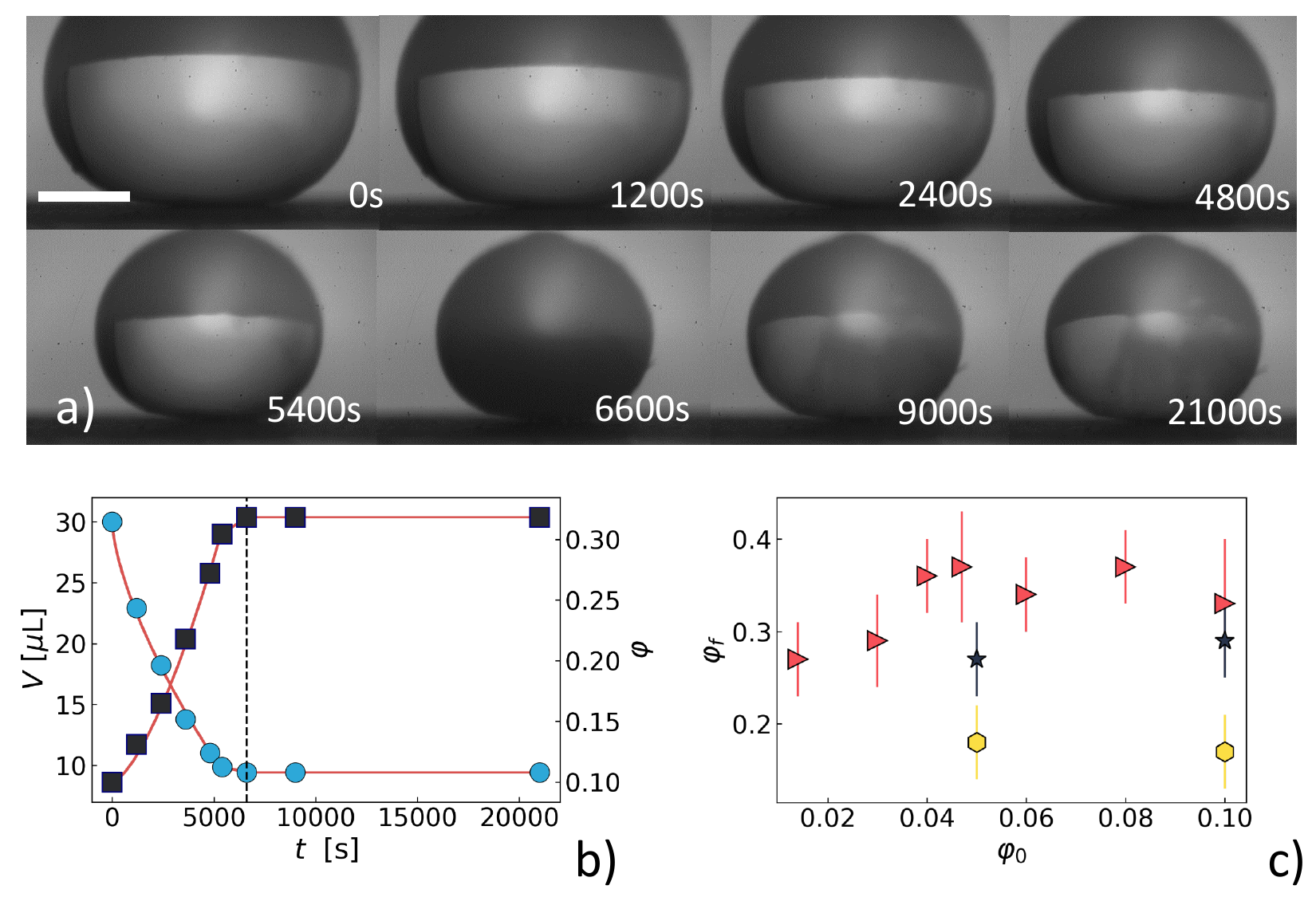}
\caption{Side view imaging of a bead of colloidal gel during drying. The gel is prepared with an initial volume fraction $\varphi_0=10\%$ (with NPs of diameter $a=6$ nm). The scale bar corresponds to $1$ mm. \textcolor{myc}{At $0$ s the bead that is deposited on the top of an hydrophobic surface has a darker upper part due to optical effect. Up to $t=6600$ s the bead shrinks homogeneously.} At $t=6600$ s, the white reflection in the center of the bead is much reduced, indicating that the bead is more opaque \textcolor{myc}{due to the air invading the gel bead. After $t=6600$ s the bead does not shrink anymore, while it progressively changes color due to the air-water exchange.} b) Left axis and blue circles: Volume of the bead, computed by standard image analysis as a function of time. The bead is deposited on the hydrophobic surface at time $t=0$. The black squares and right axis show the corresponding increase of the particle volume fraction due to drying. The red lines are guides for the eyes. The vertical line shows the beginning of the second drying regime. c) Final volume fraction as a function of initial volume fraction, different symbols represent different preparations: $a=6$ nm with water as solvent (red triangles),  $a=6$ nm and ethanol as solvent (yellow hexagons), $a=25$ nm with water as solvent (black stars). Error bars are due to  the uncertainly in measuring bead volume from the images.}
\label{Fig_imaging_drying}
\end{figure}

The minimization of the interactions between the bead and the surface during the drying process allows us to obtain crack-free beads~\cite{thiery2015water,thiery2016drying}. Moreover, we prove that the drying is isotropic, since the quasi-spherical shape is kept all along the drying stage. As an illustration, we provide in Fig.~\ref{Fig_imaging_drying} snapshots of a bead of gel prepared at $\varphi_0=10\%$ (with NPs of diameter $6$ nm) undergoing drying. During solvent evaporation the bead spontaneously undergoes a two-step process. 
\textcolor{myc}{At $t=0$ s the bead that is rinsed from silicon oil is deposited on the top of an hydrophobic surface. At this stage, due to optical effects, the bead presents a darker upper hemisphere compare to the lower one. At early time ($t<6600$ s for the sample shown in Fig.~\ref{Fig_imaging_drying}) the bead shrinks homogeneously.} At a certain time  ($t=6600$ s for the sample shown in Fig.~\ref{Fig_imaging_drying}), the volume of the bead stops decreasing. The compaction of the silica network formed by the NPs is arrested, but the gel keeps drying: the solvent evaporates, flowing through the porous silica structure, and air starts to invade the solid structure leading to some opacity in the bead (see image at $t=6600$ s in Fig.~\ref{Fig_imaging_drying}). At later times, ($t=9000$ s for the sample shown in Fig.~\ref{Fig_imaging_drying}), the bead becomes translucent, when air has fully invaded the solid structure. The two-step drying of gels is a documented process (see e.g.~\cite{brinker2013sol}), which has also been observed recently by NMR on similar gels as the ones investigated here but in bulk~\cite{coussot}. By simple mass conservation, we can compute $\varphi_f$, the final volume fraction of NPs, at the end of the shrinking process. The values of $\varphi_f$, which quantify the degree of compaction that a bead undergoes during drying, are reported in Fig.~\ref{Fig_imaging_drying}c, for various experimental conditions. For beads prepared with NPs of diameter $6$ nm, $\varphi_f = (33 \pm 4)$ \%, irrespective of the initial volume fraction $\varphi_0$, indicating that the initial volume fraction does not play a major role in the final properties of the silica bead. Beads for which water has been exchanged with ethanol stop shrinking earlier, leading to a significantly lower final volume fraction, $\varphi_f = (17.5 \pm 0.5) $ \%, as compared to the same bead with water as solvent. Similarly, beads prepared with larger NPs (diameter $25$ nm)  stop shrinking earlier than those prepared with smaller NPs, yielding also a lower final volume fraction ($\varphi_f = (28 \pm 1) $ \%). To a first order approximation, one can consider that a bead stops shrinking when it becomes energetically more favorable to allow air to invade the structure than to compress further the colloidal gel. In other words, the cross-over between the shrinking regime and the air invasion regime would be expected to occur when the Laplace pressure $\gamma/r_p$ becomes comparable to the gel Young modulus\cite{brinker2013sol}.  Laplace pressure decreases when water replaces ethanol (as the surface tension $\gamma$ decreases) or when the NPs are larger (as $r_p$ the pore size increases), thus yielding a cross-over that occurs earlier, hence a lower final volume fraction, as observed experimentally.

\subsubsection{Mechanical properties}

\begin{table}[h!]
  \centering
  \begin{tabular}{|c|c|c|c|}
    
    \hline
    $a$ (nm) &$\varphi_0$ ($\%$)&$\varphi_f($\%$)$ & $E$ (GPa) \\
    \hline
    6&$3 $ & 29 &  1.78 $\pm$ 0.06\\
    6&$4 $ & 36 &  1.72 $\pm$ 0.07\\
    6&$6 $ & 34 & 1.43  $\pm$ 0.05  \\
    6&$8 $ & 37  & 1.99 $\pm$ 0.05\\
    6&$10 $ & 33 & 1.34  $\pm$ 0.04 \\
    \hline
    25&$5 $   &27& 0.12 $\pm$ 0.06  \\
    25&$10 $ &29& 0.09 $\pm$ 0.05   \\
    \hline

  \end{tabular}
  \caption{Young modulus $E$ of mesoporous silica beads prepared in different conditions, with water as solvent. $a$ is the NP diameter, $\varphi_0$ is the initial NP volume fraction of the colloidal gel, and $\varphi_f$ is the volume fraction reached at the end of the drying process.}
  \label{tab:mec_p}
\end{table}

We measure the linear elastic Young modulus $E$ of the silica beads using a compression test. The numerical values of $E$ are reported in Tab.\ref{tab:mec_p} for the beads prepared with different experimental conditions. We find that $E$ is independent of the initial volume fraction of the hydrogels, $\varphi_0$, as anticipated since beads with various $\varphi_0$ all lead to dried beads with comparable silica volume fractions (Fig.~\ref{Fig_imaging_drying}c). The beads prepared with NPs of diameter $a=6$ nm lead to ultra stiff porous beads, with a Young modulus $E \approx1.75$ GPa, hence roughly one order of magnitude lower than the one of bulk silica. Interestingly, the beads prepared with larger NPs ($a=25$ nm) are significantly softer ($E \approx0.1$ GPa) than the ones prepared with NPs of diameter $6$ nm, as anticipated for the lower silica content of the beads beads made of larger NPs.

\subsubsection{Textural properties}

\begin{figure}
\includegraphics[width=0.5\columnwidth]{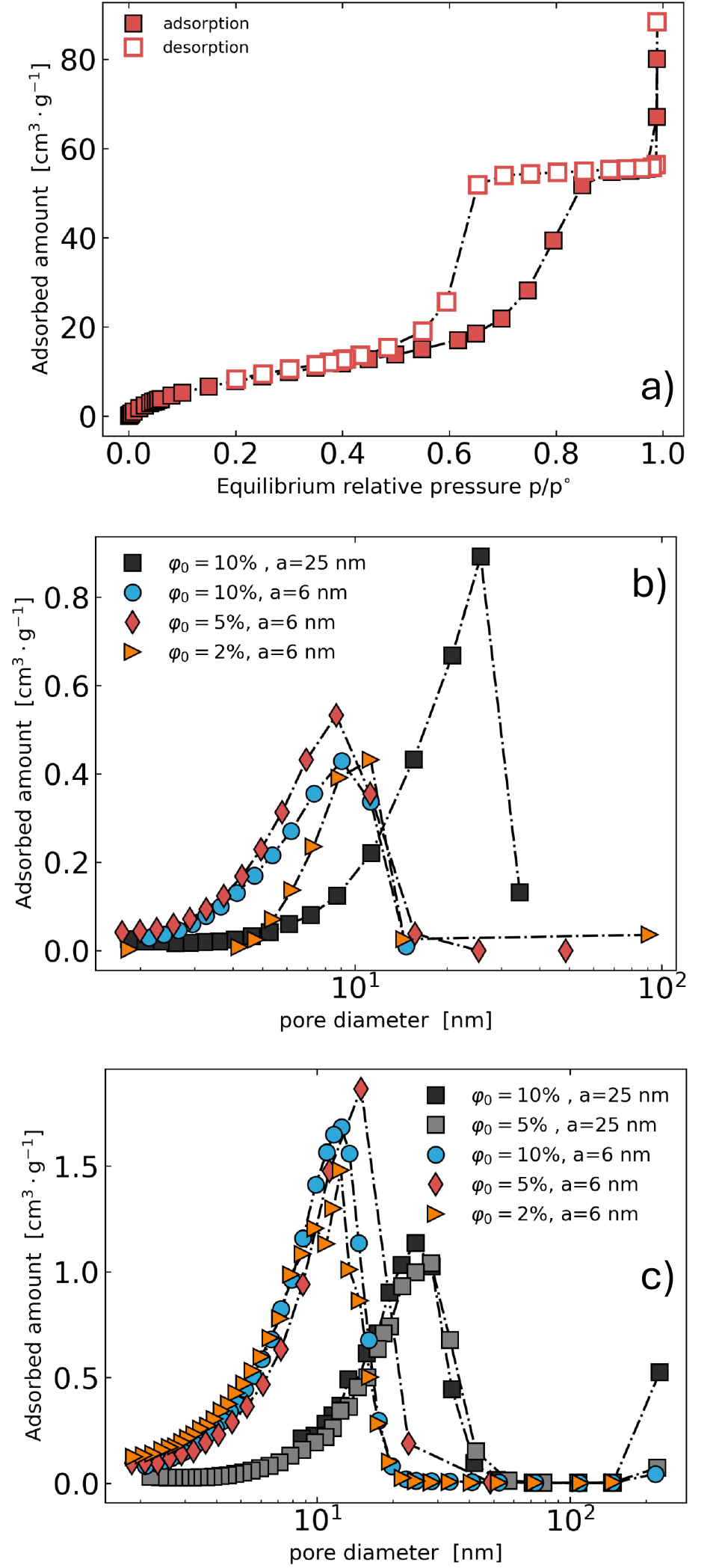}
\caption{a) Nitrogen adsorption isotherm performed at $77$ K. Full, respectively empty, symbols correspond to adsorption, respectively desorption,  data. The sample is a dried porous silica bead with $\varphi_f=35\%$ obtained from an aqueous gel prepared at $\varphi_0=10\%$, with NPs of diameter $6$ nm. b) Pore size distribution obtained using the BJH model applied to the adsorption branch of the sorption isotherm. Different symbols correspond to different samples as indicated in the legend. Silica beads have been produced by drying beads of gel with water as solvent. c)  Same plot as b) for beads that have been immersed in ethanol to perform solvent exchange before the drying step.}
\label{Fig_BET}
\end{figure}

Figure~\ref{Fig_BET}a shows a representative sorption isotherm for dried silica beads prepared at initial volume fraction $\varphi_0=10\%$ with NPs of diameter $6$ nm and with water as solvent before drying.
The shape of the sorption isotherm is classical for mesoporous materials (type IV according to the IUPAC classification), with a clear uptake at intermediate relative pressures, followed by a very flat plateau. From the monolayer-multilayer sorption mechanism ($0.05<p/p^{\circ}<0.25$) the specific surface area (SSA) could be evaluated, using the BET model~\cite{rouquerol2013adsorption}, taking $0.162$ nm$^{2}$ as cross-sectional area for nitrogen. The numerical values for SSA of all samples are reported in Tab.~\ref{tab:my_table}. We find that, for beads prepared with NPs of diameter $6$ nm, the SSA is on average around $110$ m$^{2}$.g$^{-1}$, while for beads prepared with NPs of diameter $25$ nm, we find a specific area of $81$ m$^{2}$.g$^{-1}$. Beads for which water has been exchanged with ethanol present a higher specific area: the SSA is around $310$ m$^{2}$.g$^{-1}$, for NPs of diameter $6$ nm, and  around $115$ m$^{2}$.g$^{-1}$, for NPs of diameter $25$ nm. For all samples, no macroporosity could be evidenced, as shown by the vertical slope at $p/p^{\circ}$ very close to $1$ on the sorption isotherm, which is the indication of the liquefaction of nitrogen in the sample container.

The mesoporosity can be quantified using the BJH model on the adsorption branch. The derivation is performed from the saturation plateau at $p/p^{\circ}=0.95$ to ensure a full filling of the mesopores. Figures~\ref{Fig_BET}b and c display the pore size distribution for the different experimental conditions and the numerical values for SSA and for the pore diameter at the peak of the distribution are reported in Tab.~\ref{tab:my_table}. For beads prepared with small NPs ($a=6$ nm), a rather narrow population of mesopores with diameters ranging between $4$ and $15$ nm is measured. Consistently with the results for $\varphi_f$ and the Young modulus, equivalent pore size distributions are measured for gel beads prepared with the same NPs but starting from different initial NPs volume fraction $\varphi_0$. By contrast, the pore size distribution is shifted toward larger values for a gel prepared with larger NPs, with diameter $a=25$ nm. When water is exchanged with ethanol before the drying step, the silica beads present a pore size distribution shifted toward larger values by $2$ to $3$ nm compared to the beads with water as solvent. Overall, we find that the pore size distribution is not impacted by the initial NPs volume fraction of the colloidal gel bead and is centered around a value comparable to the NPs diameter.  These findings are  consistent with the structuration of the material, the aggregation of larger particles leading to larger interparticular voids. \textcolor{myc}{Moreover, we note the absence of microporores, which could be due to local condensation between adjacent NPs, leading to progressive stiffening of solid-solid contacts, as previously postulated~\cite{manley_time-dependent_2005} and recently measured between silica particles~\cite{bonacci_contact_2020}.}

\begin{table}[h!]
  \centering
  \begin{tabular}{|c|c|c|c|c|}
    
    \hline
    a (nm) &$\varphi_0$  & solvent& SSA (m$^{2}$g$^{-1}$)& $Dp$ (nm)\\
    \hline
    6&$2 $ $\%$ &water  & $72$  & $10.0$  \\
    6&$5 $ $\%$  &water& $113$  & $8.6$ \\
    6&$10 $ $\%$  &water& $146$  & $9.0$  \\
    \hline
    25&$10 $ $\%$  &water& 81  & 25  \\
    \hline
    6&$2 $ $\%$ &ethanol  &  $312$ &  $12.3$  \\
    6&$5 $ $\%$  &ethanol&  $316$ &  $15.2$  \\
    6&$10 $ $\%$  &ethanol&  $302$ &  $12.2$  \\
    
    \hline
    25&$5 $ $\%$ &ethanol  & $125$ & $25$  \\
    25&$10 $ $\%$  &ethanol & $103$ & $28$  \\
    \hline

  \end{tabular}
  \caption{Textural properties of silica beads prepared in different conditions. $a$ is the NP diameter, $\varphi_0$ the NP volume fraction of the colloidal gel, $\varphi_f$ the NP volume fraction reached at the end of the drying process, SSA stands for specific surface area, and $Dp$ is the pore diameter at the peak of the pore size distribution}
  \label{tab:my_table}
\end{table}

\subsection{Production of CO$_2$ traps beads}

\begin{figure}
\includegraphics[width=\columnwidth]{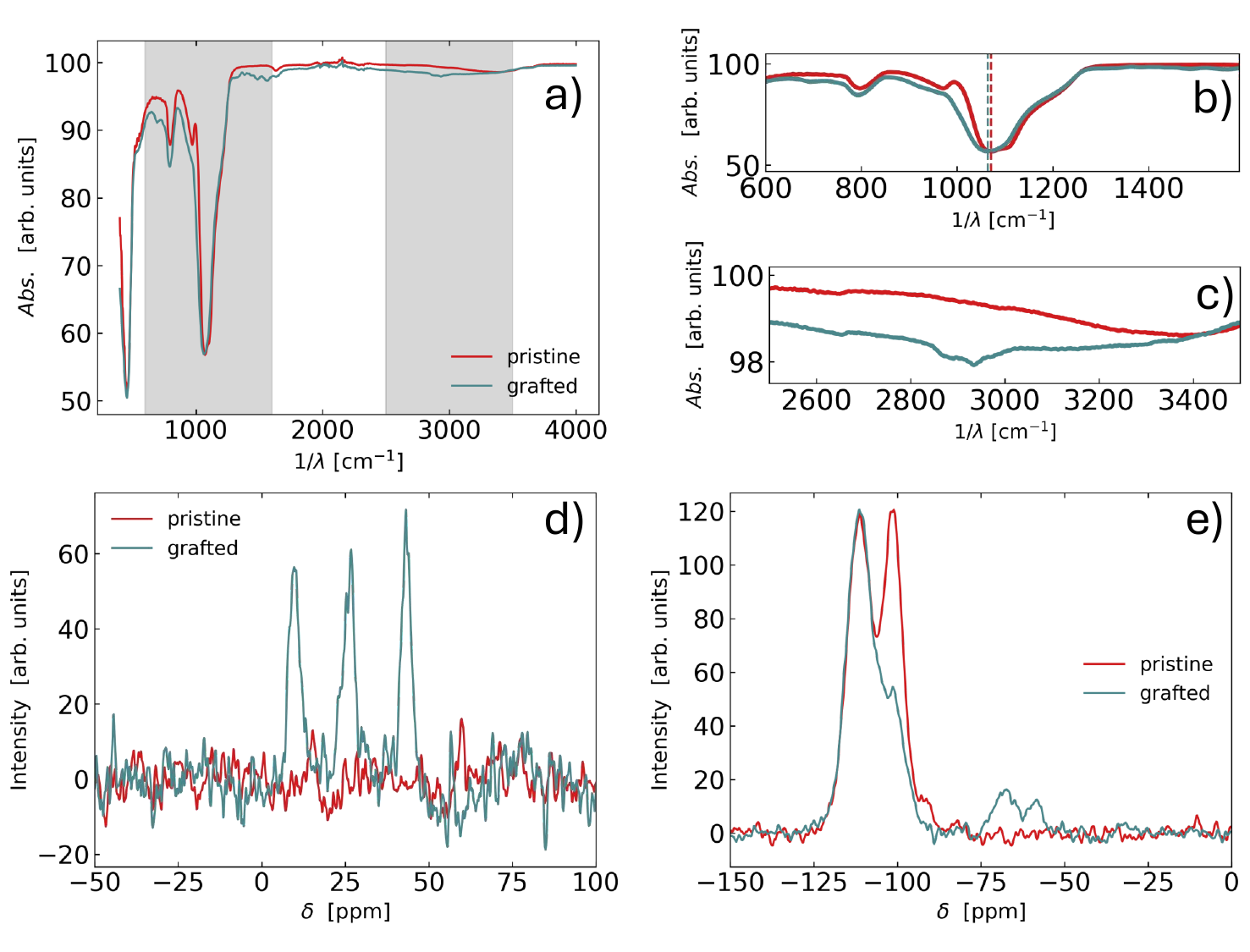}
\caption{a,b,c) Infrared spectroscopy, d) $^{13}$C CP MAS NMR,  e) $^{29}$Si CP MAS NMR, of pristine dried silica beads (red lines and symbols) and grafted beads (green lines and symbols). In a) the shaded areas indicate the zooms in the range $(2500-3500)$ cm$^{-1}$ and in the range $(600-1590)$ cm$^{-1}$ of panels b) and c) respectively. In b) the dotted lines indicate the shift of the position of the silica transmittance band from $1071$ (red) to $1064$ cm$^{-1}$ (green). }
\label{Fig_grafting}
\end{figure}

To illustrate the potentiality of the mesoporous silica beads for applications, we graft the beads with APTES to render them hydrophobic and further investigate their capability to capture carbon dioxide. For this test, we use dried beads prepared from a gel with initial volume fraction $\varphi_0=5$ \% with NPs of diameter $6$ nm. The grafting efficiency of the amino groups on the surface of the beads is assessed by comparing pristine (before grafting) and grafted beads by Fourier-transform infrared spectroscopy (FTIR) and  magic-angle spinning (MAS) NMR (Fig.~\ref{Fig_grafting}). 
On the infrared spectrum corresponding to the pristine beads, the main typical transmittance bands characterizing pure silica can be seen at $1071$ cm$^{-1}$ but also at $790$ cm$^{-1}$. Upon grafting, these bands are still present, even though a shift from $1071$ down to $1064$ cm$^{-1}$ can be observed. On the other hand, new transmittance bands can be seen in the region between $1490$ cm$^{-1}$ and $690$ cm$^{-1}$, which correspond to the bending vibrations of -NH$_2$ and -CH$_2$ functions. The presence of APTES moieties on the silica surface can be further confirmed by the -CH$_2$ stretching in the region $(2850-2950)$ cm$^{-1}$, only visible after APTES grafting (Fig.~\ref{Fig_grafting}c).
We thus conclude that APTES molecules are located on the surface of the silica beads. However, despite a thorough washing of the beads, the covalent bonding of APTES is not certain, even though the downshift of the Si-O-Si transmittance band at $1071$ cm$^{-1}$ is an indication of the perturbation of the siloxane network. \textcolor{myc}{To have further evidence of the grafting}, solid state NMR is performed on pristine and grafted silica beads. We show in Fig.~\ref{Fig_grafting}d the $^{13}$C CP MAS NMR spectra of the beads before and after grafting. For pristine beads, no carbon signal is detected, as expected since the beads have been calcined. After grafting with APTES, three signals can be detected that correspond to the propyl fragments of the APTES molecule. The signal at $\sim10$ ppm can be attributed to the CH$_2$ group attached to the Si atom while the signal at $\sim25$ ppm can be attributed to the CH$_2$ group in $\beta$ position with the Si atom. Finally, the signal at $\sim45$ ppm can be attributed to the CH$_2$ group attached to the amino function. The chemical shift is classically attributed to the higher unshielding of the CH$_2$ groups due to the amino function compared to the Si function.
These three signals cannot be attributed to the presence of hydrolysed APTES as APTES is soluble in the solvents used during the synthesis. If present, it would have been washed during the preparation of the beads. We therefore conclude that APTES molecules are covalently grafted on the surface of the beads. To further confirm this result, we perform $^{29}$Si solid state NMR (Fig.~\ref{Fig_grafting}e).
The $^{29}$Si NMR spectrum of pristine beads exhibits two main signals that are typical of condensed silica. At $\sim-115$ ppm, the NMR signal corresponds to Q$^3$ (isolated silanols) and Q$^4$ environments (condensed silica). Hence, the pristine beads are made of condensed silica with some isolated silanols. At $\sim-100$ ppm, a signal typical of a Q$^2$ environment is detected, which corresponds to a silica surface bearing geminated silanols. The signal at $-92$ ppm, which can be attributed to Q$^1$ environments (Si(-O)$_3$), is weak, suggesting that Si(-O)$_3$ are very limited in the pristine silica beads.
When looking at the grafted beads spectrum, the Q$^2$ peak almost vanishes while the Q$^3$ and Q$^4$ peaks are still clearly present. Hence, we conclude that silica is more condensed after APTES grafting. Indeed, surface silanols (Si-OH) are replaced with Si-O-Si(APTES) leading to an increase of the condensation. At $-70$ ppm, a clear signal can be seen. It corresponds to T$^3$ environments on the grafting agent side. Si(-O)$_3$ are therefore visible, which suggests that APTES can be attached to the surface through three Si-O-Si bonds. However, T$^2$ environments can be noticed at $-60$ ppm, which implies that some APTES moieties are not completely condensed, with one pending silanol per APTES molecule.
Overall, infrared and NMR results suggest that APTES is strongly grafted on the silica surface of the beads. This could be further evidenced by adsorbing a specific probe such as CO$_2$.

\begin{figure}
\includegraphics[width=0.7\columnwidth]{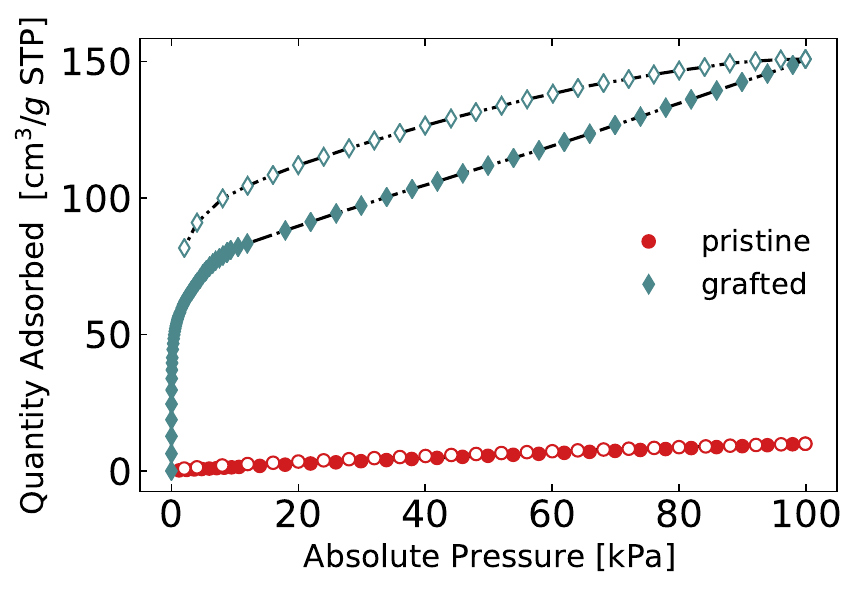}
\caption{CO$_2$ adsorption of pristine (red circles) and grafted (green diamonds) mesoporous silica beads. The dried beads are prepared from a gel with initial volume fraction $\varphi_0=5\%$. Full, respectively empty,  symbols correspond to the sorption, respectively desorption, curves.}
\label{Fig_CO2}
\end{figure}

Figure~\ref{Fig_CO2} displays the carbon dioxide sorption isotherms for both pristine and grafted beads.
The isotherm obtained with the pristine beads reveals a very low CO$_2$ sorption capacity, which is consistent with the expected weak interaction at $298$ K between CO$_2$ and the surface silanols of the silica. In sharp contrast, in the case of grafted beads, the sorption capacity is significantly increased, from $10$ up to $150$ cm$^{3}\cdot$g$^{-1}$ at atmospheric pressure. Because the pristine material has a specific surface area twice as large as that found for the material after APTES grafting ($310$ versus $154$ m$^{2}\cdot$g$^{-1}$), this difference at high pressure cannot be attributed to the textural properties of the materials. Therefore, the APTES functions are responsible for the CO$_2$ sorption. More importantly, the Henry’s constants are very different for pristine ($1.6 \times 10^{-4}$ cm$^{3}$$\cdot$g$^{-1}$ Pa$^{-1}$) and grafted ($1.5$ cm$^{3}$$\cdot$g$^{-1}$ Pa$^{-1}$) beads, ensuring that the affinity of CO$_2$ molecules for the grafted beads is  higher, thus confirming that APTES functions are efficiently grafted on the silica surface. The large hysteresis loop observed for the grafted beads is also indicative of the strong interaction between CO$_2$ and the grafted beads.

\section{Conclusion}

This work presented a simple two-step protocol to produce stiff spherical mesoporous silica beads of millimetric size. The first step consisted in producing beads of colloidal gel of controlled size and microstructure, by exploiting an in-situ gelation reaction that ensured a spatially uniform microstructure. To vary the properties of the beads, and thereafter the textural properties of the dried beads, we varied the initial concentration of silica nanoparticles (NPs) and their size, with an optional step consisting in the solvent exchange with ethanol once the in-situ gelation was complete.
Once the beads of colloidal gels were formed, they were dried in ambient conditions to produce dried porous silica beads. The spherical shape of the beads and the hydrophobic surface over which they were deposited minimized the tensile stresses that often raise during the drying process, which allowed us to obtain crack-free stiff silica beads. Depending on the experimental conditions, the silica volume fraction of the beads lied in the range $(0.18-0.38)$ (corresponding to a porosity between $62$ and $82$ \%), while their Young modulus varied between $0.09$ and $2$ GPa. Using adsorption techniques, we showed that the beads were mesoporous, with a narrow pore size distribution peaked at $10$ or $25$ nm depending on the NP size, and a specific surface up to more than $300$ m$^{2}$.g$^{-1}$, when the colloidal gel was prepared with NPs of diameter $6$ nm and using the water-ethanol solvent exchange.
Finally, to further show the potential of our mesoporous silica beads, we successfully grafted APTES molecules onto their surface. We demonstrated that the grafted beads act as CO$_2$ traps, making them promising candidates for storage and adsorption of gases, catalysis, and chromatography applications.

\begin{acknowledgement}

We thank J. Barbat for help in designing and building the DLS environment, N. Shafquat for help in TGA measurements, T. Bizien for the X-ray measurements, J.-M. Fromental for help with the rheological measurements, and W.C.K. Poon for enlightening discussions. We acknowledge financial support from the French Agence Nationale de la Recherche (ANR) (Grant No. ANR-19-CE06-0030-02, BOGUS), and from the Synchrotron Soleil. LC gratefully acknowledges support from the Institut Universitaire de France.

\end{acknowledgement}


\bibliography{achemso-demo}

\end{document}